\documentclass[%
 reprint,
superscriptaddress,
 amsmath,amssymb,
 aps,pra,twocolumn
]{revtex4-1}

\usepackage{graphicx}
\usepackage{dcolumn}
\usepackage{bm}
\usepackage{algcompatible}

\newcommand{\vect}[1]{\ensuremath{\boldsymbol{#1}}}
\newcommand{\E}[1]{\times 10^{#1}}

\usepackage{color,hyperref}
\definecolor{linkColor}{rgb}{1,0,0}
\hypersetup{pdfborder={0 0 0},colorlinks=true,urlcolor=linkColor,citecolor=linkColor}

\emergencystretch=\maxdimen
\begin{document}

\preprint{APS/123-QED}

\title{A linear-scaling algorithm for rapid computation of inelastic transitions in the presence of multiple electron scattering}

\author{Hamish G. Brown}
 \affiliation{ National Center for Electron Microscopy, Molecular Foundry, Lawrence Berkeley National Laboratory, California 94720, USA}
  \email{hamishbrown@lbl.gov}
\author{Jim Ciston}%
 \email{jciston@lbl.gov}
\affiliation{ National Center for Electron Microscopy, Molecular Foundry, Lawrence Berkeley National Laboratory, California 94720, USA}
\author{Colin Ophus}%
 \email{clophus@lbl.gov}
\affiliation{ National Center for Electron Microscopy, Molecular Foundry, Lawrence Berkeley National Laboratory, California 94720, USA}

\date{\today}

\begin{abstract}
Strong multiple scattering of the probe in scanning transmission electron microscopy (STEM) means image simulations are usually required for quantitative interpretation and analysis of elemental maps produced by electron energy-loss spectroscopy (EELS).  
These simulations require a full quantum-mechanical treatment of multiple scattering of the electron beam, both before and after a core-level inelastic transition. 
Current algorithms scale quadratically and can take up to a week to calculate on desktop machines even for simple crystal unit cells and do not scale well to the nano-scale heterogeneous systems that are often of interest to materials science researchers. 
We introduce an algorithm with linear scaling that typically results in an order of magnitude reduction in compute time for these calculations without introducing additional error and discuss approximations that further improve computational scaling for larger scale objects with modest penalties in calculation error.
We demonstrate these speed-ups by calculating the atomic resolution STEM-EELS map using the L-edge transition of Fe, for of a nanoparticle 80 \AA\ in diameter in 16 hours, a calculation that would have taken at least 80 days using a conventional multislice approach.
\end{abstract}


\maketitle

\section{Introduction}

The presence of even small numbers of dopant atoms in a material can have a disproportionate effect on that material's properties, and thus characterization techniques that perform nanoscale chemical and elemental mapping are important investigative tools in materials science. 
Furthermore, the presence of extended, non-periodic defects, interfaces, and nanoscale precipitate phases either naturally occurring or engineered for functionality requires quantitative atomic resolution analysis of fields of view that are substantially larger than crystal unit cells~\cite{xin2011atomic,mundy2012atomic,mundy2014visualizing,zamani2017atomic}.
In scanning transmission electron microscopy electron energy-loss spectroscopy (STEM-EELS) a focused probe, which can be smaller than the width of an atom in state-of-the-art instruments, is raster scanned across a material, and the energy-loss spectrum is recorded for each scan position of the probe.
Integrating regions of the spectra that correspond to energy-losses characteristic of ionization  of a particular atomic element allows for mapping the positions and concentrations of elements within the specimen~\cite{egerton2011electron}.
Additionally, analysis of the fine structure of the energy loss spectrum can reveal modified valence state of the elements due to bonding~\cite{mundy2012atomic,muller1993mapping,muller2008atomic}.

A complicating factor for STEM-EELS interpretation is that strong multiple scattering of the electron probe before and after exciting an ionization event means that for quantitative work, such as measuring chemical concentrations, STEM-EELS results often need to be interpreted by performing forward simulations of an assumed structure~\cite{bosman2007two}. 
The simulations typically combine either multislice or Bloch wave simulations with inelastic scattering cross-sections for different elements of interest~\cite{bosman2007two,allen1997inner,allen2015modelling,lobato2016progress}. 
Currently, STEM-EELS image simulations of simple crystalline structures can require up to a week of calculation time, even using graphical processing unit (GPU) accelerated simulation codes~\cite{dwyer2010simulation,dwyer2015role}. 
Algorithms with much faster run times that scale better with system size are required to simulate large objects, or those with more heterogenous chemical properties than simple crystals, such as samples containing dopants, extended defects, interfaces, or entire nanoparticles. 
Recent improvements in the readout speed of electron spectrometers adds additional impetus to efforts to speed up STEM-EELS simulations, as faster spectrometers allow for more routine imaging of much larger systems in STEM-EELS at atomic resolution.

A recent innovation in the field of electron microscope image simulation is the development of the plane wave reciprocal-space interpolated scattering matrix (PRISM) algorithm~\cite{ophus2017fast}.  
Instead of the traditional approach of calculating the propagation of the beam for each scan position in a STEM raster independently, the PRISM method uses the multislice algorithm to calculate the scattering matrix for a particular electron microscope experiment and stores it in memory.
The scattering matrix, which propagates components of the electron illumination through the imaging object, can then be rapidly applied to the illumination wave function for each scan position to compute the output wave~\cite{pryor2017streaming}. 

In this paper, we extend the PRISM algorithm to the problem of STEM-EELS image simulation, and demonstrate substantial speed up in calculation time. This paper is structured as follows: in Section~\ref{sec:theory} the underlying physics of STEM-EELS simulation are reviewed and we discuss both the scattering matrix operator and the multislice algorithm for numerical simulation of STEM images. 
In Section~\ref{sec:algorithms} we detail algorithmic implementation of both the conventional multislice approach and our novel PRISM approach to STEM-EELS calculation and discuss the predicted scaling of the run-time of both algorithms with different parameters such as simulation grid size and specimen thickness.
Section~\ref{sec:Implementation} shows results simulated using the new method, and compares accuracy and speed-up relative to the conventional multislice approach.
Two further optimizations are discussed in this section: evaluating inelastic transitions on a cropped grid and using an inverse multislice operation to economize on the total number of multislice operations.
These optimizations are shown to typically induce minimal errors while offering substantial reductions in compute time.
This section is concluded with a calculation of the STEM-EELS image for an illustrative large heterogeneous object, in this case a FePT nanoparticle roughly 80 \AA\ in diameter~\cite{yang2017deciphering}.
%
%
%
\section{Theory \label{sec:theory}}
In this section we briefly review the underlying physical theories behind the simulation of a relativistic electron propagating through a specimen of condensed matter.
We introduce the transition potential to simulate the inelastic scattering of the probe electron due to the ionization of an electron bound to an atom within the specimen. 
Existing methods calculate the propagation of the electron probe to the plane of ionization and subsequent propagation of each inelastically scattered wave separately for each scan position and each inelastic transition of interest. 
We outline how we can calculate and store the scattering matrix operator that performs all of these steps and then rapidly apply it to all probes in the STEM raster.  

We begin with the  Schr\"odinger equation in reciprocal space for a fast electron interacting with the electrostatic potential of a specimen of condensed matter~\cite{van1985image},
\begin{align}
\frac{d \hat{\psi}_{\vect{g}}(z)}{dz} =& -i\pi\lambda g^2\hat{\psi}_{\vect{g}}(z) + \sum_{\vect{h}} i\sigma \hat{V}_{\vect{g-h}}\hat{\psi}_{\vect{h}}(z)\,.
\label{eq:Schro}
\end{align}
The $\hat{\psi}_{\vect{g}}(z)$ are Fourier coefficients of the fast electron wave function as a function of depth $z$ in the specimen and Fourier space coordinates $\vect{g}$ (magnitude $g$) and $\vect{h}$ in the plane perpendicular to the direction of propagation. 
The Fourier coefficients of the electrostatic potential are denoted by $\hat{V}_{\vect{g}}$. 
The interaction constant $\sigma = 2\pi m_e e\lambda/h^2$, where $m_e$ and $\lambda$ are the (relativistically corrected) mass and wavelength of the electron, $e$ is the electron charge and $h$ is Planck's constant. Equation ($\ref{eq:Schro}$) is a set of coupled linear equations for which the solution can be written as the matrix-vector product
\begin{align}
\boldsymbol{\psi}(z) = \sum_{\vect{h}}e^{-i\pi\lambda zg^2\delta_{\vect{gh}}+i\sigma \hat{V}_{\vect{g-h}}z}\hat{\psi}_{\vect{h}}(0) \equiv \mathcal{S}(z)\boldsymbol{\psi}(0) \label{eq:Spsi}
\end{align}
where  $\delta_{\vect{gh}}$ is the Kronecker delta and the bold $\boldsymbol{\psi}$ is a vector containing the Fourier coefficients of the illumination $\hat{\psi}_{\vect{h}}$. For STEM the illumination is a coherent focused probe with functional form
\begin{align} \label{eq:illum}
    \hat{\psi}(\vect{g},0) = A(\vect{g})e^{-i\chi(\vect{g})}\,.
\end{align}
Here $A(\vect{g})$ is the aperture function, a top-hat function with radius equal to the convergence semi-angle of the probe, and $\chi(\vect{g})$ is the aberration function which takes into account probe aberrations such as defocus, spherical aberration and astigmatism~\cite{kirkland2010advanced}.

Efficient numerical calculation of Eq.~(\ref{eq:Spsi}) typically proceeds through diagonalisation of the scattering matrix $\mathcal{S}$ (the Bloch wave method~\cite{bethe1928theorie,Humphreys}), or through a split-step evaluation of the action of $\mathcal{S}$ on $\vect{\psi}$, with the specimen first split into $n$ slices in the beam direction and the operator involving the propagation matrix elements $-\pi\lambda g^2\delta_{\vect{gh}}z/n_z$ and that involving the specimen interaction matrix elements $\sigma \hat{V}_{\vect{g-h}}z/n_z$ applied in alternating sequence: 
\begin{align}
\boldsymbol{\psi}(z) &= \left[e^{-i\pi\lambda zg^2\delta_{\vect{gh}}/n_z+i\sigma \hat{V}_{\vect{g-h}}z/n_z}\right]^{n_z}\boldsymbol{\psi}(0)\,, \nonumber  \\
&\approx\left[e^{-i\pi\lambda zg^2\delta_{\vect{gh}}/n_z}e^{i\sigma \hat{V}_{\vect{g-h}}z/n_z}\right]^{n_z}\boldsymbol{\psi}(0)\,. \label{eq:3}
\end{align}
This is called the ``multislice method'' in the electron microscopy literature \cite{van1985image} and has become the most popular method of evaluation of Eq.~(\ref{eq:Spsi}). 
This is because the $e^{i\sigma \hat{V}_{\vect{g-h}}z/n_z}$ operator (referred to as the transmission function) is diagonal in real space and a fast Fourier transform (an FFT, which we shall represent with the symbol $\hat{\mathcal{F}}_{\vect{\mathrm{r}\rightarrow\mathrm{g}}}$ and it's inverse operation as $\hat{\mathcal{F}^{-1}}_{\vect{\mathrm{g}\rightarrow\mathrm{r}}}$)\footnote{We use the following convention, a forward  Fourier transform is given by $\hat{\psi}(\vect{g}) = \hat{\mathcal{F}}_{\vect{\vect{r}\rightarrow\vect{g}}}\psi(\vect{r}) = \int \psi(\vect{r})e^{-2\pi i \vect{r}\cdot\vect{g}}d\vect{r}$ and an inverse Fourier transform is given by $\psi(\vect{r}) = \hat{\mathcal{F}}^{-1}_{\vect{g}\rightarrow\vect{r}}\hat{\psi}(\vect{r}) = \int \hat{\psi}(\vect{g})e^{2\pi i \vect{r}\cdot\vect{g}}d\vect{r}$} can be used to efficiently transform between real and reciprocal space,
\begin{align}
\boldsymbol{\psi}(z) \approx \left[e^{-i\pi\lambda z/n_zg^2\delta_{\vect{gh}}}\hat{\mathcal{F}}_{\vect{\mathrm{r}\rightarrow\mathrm{h}}}e^{i\sigma V(\vect{r})z/n_z}\hat{\mathcal{F}}^{-1}_{\vect{\mathrm{g}\rightarrow\mathrm{r}}}\right]^{n_z}\boldsymbol{\psi}(0)\
\,. \label{eq:4}    
\end{align}
The two-dimensional FFT has a favourable $2N^2\log N$ scaling, where $N$ is  the number of Fourier coefficients $\vect{g}$ included in the simulation. 
For brevity we will define the following operator as short hand for a single multislice iteration,
\begin{align} \label{eq:MSop}
   \mathcal{M}(\Delta z)\equiv e^{-i\pi\lambda z / n_z g^2 \delta_{\vect{gh}}} \hat{\mathcal{F}}_{\vect{\mathrm{r}\rightarrow\mathrm{h}}}e^{i\sigma V(\vect{r})z/n_z}\hat{\mathcal{F}}^{-1}_{\vect{\mathrm{g}\rightarrow\mathrm{r}}}\,.
\end{align}
By comparison the Bloch wave approach, which solves Eq.~(\ref{eq:Spsi}) through diagonalisation of a matrix operator containing the coefficients within the exponent of Eq.~(\ref{eq:Spsi}), has the less favourable $N^3$ scaling~\cite{findlay2003lattice}. 
A more in-depth explanation of the multislice algorithm with details on it's implementation may be found in Ref.~\cite{kirkland2010advanced}. 

A recent innovation in STEM image simulation is the PRISM (plane-wave reciprocal-space interpolated scattering matrix) algorithm, in which only the rows of the matrix operator $\mathcal{S}(z)$, from Eq.~(\ref{eq:Spsi}), corresponding to the Fourier coefficients present in the illumination vector $\psi_{\vect{g}}$ are calculated using the multislice algorithm Eq.~(\ref{eq:4}) and stored in memory. 
As will be explained in the next section, this approach can be used to accelerate the calculation of STEM-EELS images because the wavefunctions for the inelastically scattered electron can be rapidly propagated to the exit surface with the stored operator.
This is instead of having to perform the multislice operation separately for each seperate probe position and inelastic scattering event in the conventional multislice approach.
\begin{figure*}[t]
    \centering
    \includegraphics[width=\textwidth]{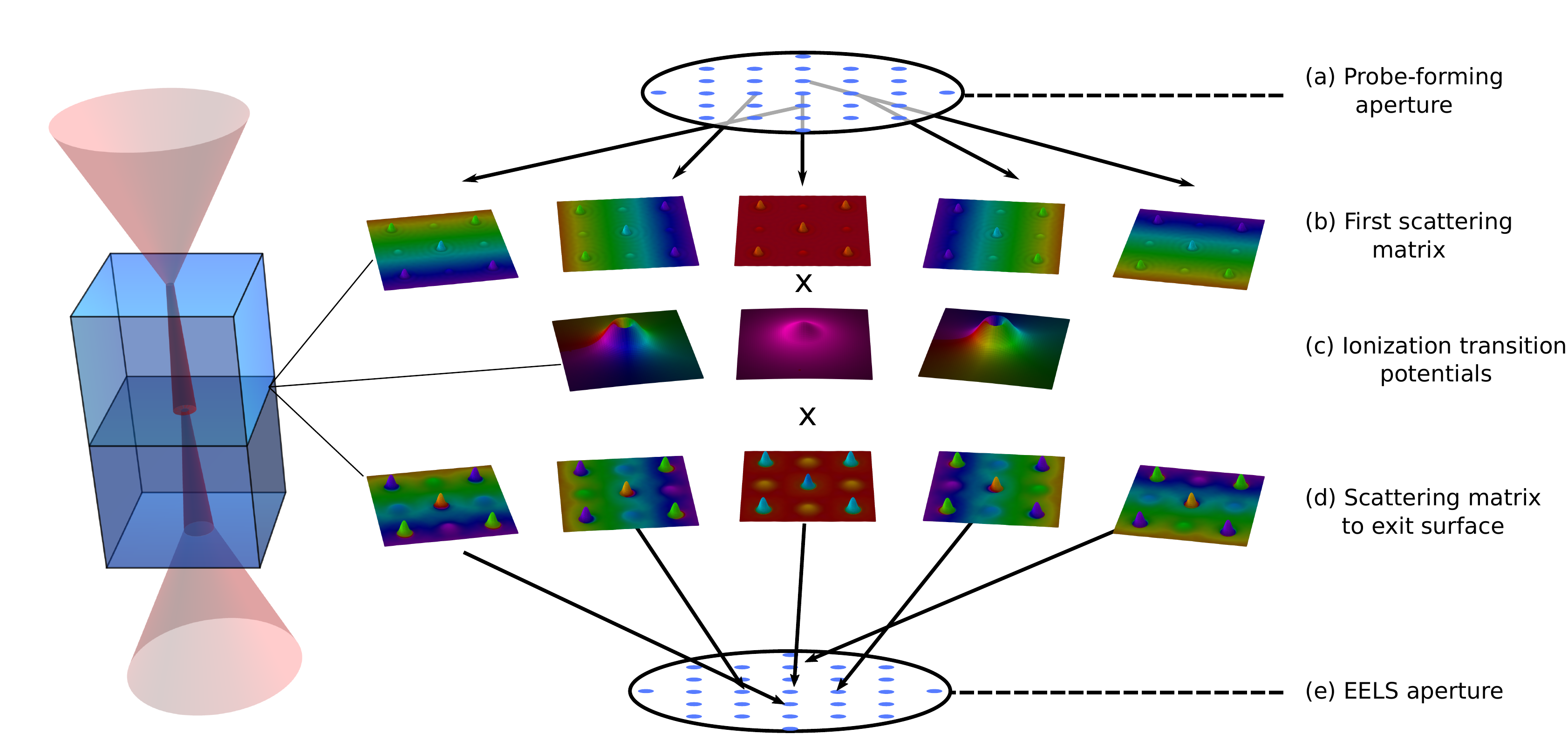}
    \caption{A diagramatic representation of the PRISM algorithm for STEM-EELS simulation that is encapsulated in Eq.~(\ref{eq:Hn0op}). The first scattering matrix maps points in the probe-forming aperture (a) to complex output in the plane of the ionization (b). At this plane the output is multiplied by the ionization transition potentials of interest $H_{n0}$ which are shown in (c) and then by the output of the second scattering matrix (d) which propagates the resulting inelastically scattered wave to reciprocal space points inside the EELS aperture (e).}
    \label{fig:1}
\end{figure*}

For an inelastic transition at depth $z$, the scattered wave function $\psi_n(\mathrm{r},z)$, where $n$ is shorthand for an excited quantum state, is given by the product of the elastic wavefunction $\psi_0(\mathrm{r},z)$ and an inelastic transition potential for ionization $H_{n0}(\mathrm{r},z)$,
\begin{align}
    \psi_n(\vect{r}_\perp,z) = H_{n0}(\vect{r}_\perp)\psi_0(\vect{r}_\perp,z)\,. \label{eq:Hn0Psi}
\end{align}
The inelastic transition potential is given by Ref.~\cite{saldin1987theory},
\begin{equation}
    H_{n0}(\vect{r}) = 
        \int u_n^*(\boldsymbol{\kappa},\vect{r}')
        \cfrac{e^2}{ 4\pi \epsilon_0  
        \lvert \vect{r} - \vect{r^\prime } \rvert } u_0(\vect{r}')d\vect{r}'
    \label{eq:Hn0def}
\end{equation}
where $u_0(\vect{r}')$ is the wave function of an electron in a bound state of the specimen and $u_n(\boldsymbol{\kappa},\vect{r}')$ is the wavefunction of that electron excited to a continuum (free) state with wave vector $\boldsymbol{\kappa}$.
The wave function in the continuum state is typically expanded in a spherical harmonic basis,
\begin{align} \nonumber
    u_n(\boldsymbol{\kappa},\vect{r}) = \frac{4\pi}{2\kappa r}&\sum_{\ell^\prime=0}^{\infty} i^{\ell^\prime}e^{i\delta_{\ell^\prime}}u_{\kappa \ell^\prime}(r)\\
    &\times\sum_{m_{l^\prime}=-\ell^\prime}^{\ell^\prime}Y_{\ell^\prime,m_{\ell^\prime}}^*(\hat{\boldsymbol{\kappa}})Y_{\ell^\prime,m_{\ell^\prime}}(\hat{\vect{r}})\,
\end{align}
where the $Y_{\ell^\prime,m_{\ell^\prime}}$ are spherical harmonic functions, $\ell$, $m_{\ell}$, $\ell^\prime$ and $m_{\ell^\prime}$ are the angular momentum and azimuthal angular momentum quantum numbers of the bound and continuum states and $\delta_{\ell^\prime}$ is a phase factor determined by fitting to an asymptotic form of the free state~\cite{saldin1987theory,oxley1998delocalization}.
To a good approximation, only a small number of the different possible $(\ell^\prime,m_{\ell^\prime})$ need be included in a converged calculation~\cite{dwyer2015role}.
The kernel in Eq.~(\ref{eq:Hn0def}) is the Coulomb potential, which mediates interactions between the fast electron and the sample, and $\epsilon_0$ is the permittivity of free space. Calculations of Eq.~(\ref{eq:Hn0def}) used in this paper are based on a solution using an angular momenta basis for $u_0$ and $u_n$ which is derived in Ref.~\cite{saldin1987theory}, the numerical implementation of which is discussed in Ref.~\cite{oxley1998delocalization}. 

We may simulate a STEM-EELS image by then propagating the wave function for the inelastically scattered electron, $\psi_n$ from Eq.~(\ref{eq:Hn0def}), to the exit surface using Eq.~(\ref{eq:Spsi}). 
We represent this process mathematically
\begin{align} \label{eq:Hn0op}
    \boldsymbol{\psi}_{n}(\vect{r},t) &=  \mathcal{S}_1\mathcal{F} \mathcal{H}_{n0}\mathcal{F}^{-1}\mathcal{S}_2\boldsymbol{\psi}_{0}(0)\, \\  \label{eq:SHn0}
    &\equiv  \mathcal{S}_{n}\boldsymbol{\psi}_{0}(0)\,.
\end{align}
Here $\mathcal{H}_{n0}$ is a matrix containing the values of $H_{n0}(\vect{r})$ on its diagonals and the matrix operator $\mathcal{S}_{n}$ we have defined in Eq.~(\ref{eq:SHn0}) for imaging of a single inelastic transition. 
Equation (\ref{eq:Hn0op}) must be solved for each inelastic transition of interest. 
Figure~\ref{fig:1} is a diagramatic representation of Eq.~(\ref{eq:SHn0}), Fig.~\ref{fig:1}(a) shows how $\mathcal{S}_1$ maps Fourier coefficients of the probe wave function to wave functions at the depth of the inelastic transition (b) where we multiply these wave functions by the ionization transition potentials in Fig.\ref{fig:1}(c) and then propagate them using $\mathcal{S}_2$, Fig.\ref{fig:1}(d),  to points within the EELS aperture in Fig.\ref{fig:1}(e).
\section{Algorithms \label{sec:algorithms}}
In this section we outline the details of the implementation of the simulation algorithms and write down estimates for the runtimes of each algorithm.
We start with the algorithm for calculating STEM-EELS images using the multislice method:
    \begin{algorithmic}
    \STATE{Loop over probe positions}
    \FOR{$x$=$1$ to $n_x$}     
    \FOR{$y$ =$1$ to $n_y$}
\STATE{Initialise illumination wave function \\ $\psi(\vect{g},0) = A(\vect{g})e^{-2\pi i (x,y)\cdot\vect{k}}$}
\STATE{Loop over slices of specimen}
\FOR{$i_z=1$ to $n_z-1$} 
\STATE{Loop over inelastic transitions within slice $i_z$}
\FOR{$n=1$ to $n_{\text{states},i}$} 
\STATE{Calculate inelastically scattered electron wavefunction $\psi_n(\vect{r}_\perp,iz) = H_{n0}(\vect{r}_\perp)\psi_0(\vect{r}_\perp,iz)$}
     \STATE{Propagate $\psi_n$ to exit surface of specimen}
     \FOR {$i_z'=i_z$ to $n_z$}
     \STATE{Multislice to advance $\psi_n$ one slice\\     $\psi_n(\vect{g},iz') = \mathcal{M}(\Delta z)\psi_n(\vect{h},iz'-1)$}
    \ENDFOR
    \STATE{Integrate wavefunction over detector function $D(\vect{g})$ and add contribution to STEM image}
    $I(x,y) = I(x,y) + \int|\psi_n(\vect{g},n_z)|^2D(\vect{g})d\vect{g}$
    \ENDFOR
    \STATE{Multislice to advance $\psi_0$ one slice\\     $\psi(\vect{g},iz+1) = \mathcal{M}(\Delta z)\psi(\vect{h},iz)$}
    \ENDFOR
    \ENDFOR
    \ENDFOR
    \end{algorithmic}
Assuming a constant number of inelastic transitions $n_{states}$ at each slice, which is not true in general but is a useful assumption for estimating the computational complexity of the above algorithm, this algorithm requires
\begin{align}
N_{MS}&=\sum_{i_z=0}^{n_z-1}[n_{\text{states}}(n_z-i_z)]+n_z\\
&=n_{\text{states}}n_z(n_z-1)/2+n_z  
\end{align}
multislice operations for each probe position. The arithmetic sum identity $\sum_{i=0}^{n-1}i=n(n-1)/2$ has been invoked in the above equation.

In our proposed PRISM STEM-EELS algorithm we calculate two scattering matrices at each slice, the first to propagate the probe wave function to the slice of the inelastic transition and the second to propagate the wave function for the inelastically scattered electron to the exit surface.

    \begin{algorithmic}
    \STATE{Initialize scattering matrices} \\
    \STATE{$\mathcal{S}_1 = \mathcal{I}$} \\
    \STATE{$\mathcal{S}_2 = [\mathcal{M}^{T}(\Delta z)]^{n_z}$}
    \FOR{$i_z=1$ to $n_z$}
    \STATE{Loop through transitions within slice $i_z$}
    \FOR{$n=1$ to $n_{\text{states},i_z}$}
    \STATE{Calculate Eq. (\ref{eq:SHn0})\\ $\mathcal{S}_{n,i_z} = \mathcal{S}_2\mathcal{F}\mathcal{H}_{n0}\mathcal{F}^{-1}\mathcal{S}_1$}
    \STATE{Apply $\mathcal{S}_{n,i_z}$ to each illumination vector in the raster scan and add the resulting amplitude to the STEM image}
    \STATE{$I(x,y) =I(x,y) + |\mathcal{S}_{n,i_z}\psi(\vect{g},x,y)|^2$}
    \ENDFOR
     \STATE{Advance $\mathcal{S}_1$ one slice $\mathcal{S}_1(i_{z+1}) = \mathcal{M}(\Delta z)\mathcal{S}_1(i_{z})$\\}
     \STATE{Re-calculate $\mathcal{S}_2$, $\mathcal{S}_2 = [\mathcal{M}^{T}(\Delta z)]^{n_z-i_z}\mathcal{I}$}   
    \ENDFOR
    \end{algorithmic}
An important point to note is that, as defined here, the rows of $\mathcal{S}_1$ correspond to points in reciprocal space within the probe forming aperture and the columns correspond to real space points in the specimen plane, so $\mathcal{S}_1$ is generally a non-square matrix due to the different sampling of these two planes.
Each column can be calculated by using multislice to propagate the plane wave components that fall withing the probe forming aperture through the specimen, as shown in Fig.~\ref{fig:1}~\cite{ophus2017fast}. 
Conversely, the rows of $\mathcal{S}_2$  correspond to real space points in the specimen plane and the columns correspond to points within the EELS detector. 
The most efficient way to calculate $\mathcal{S}_2$ is to propagate plane wave components with transverse momenta within the EELS acceptance aperture \emph{back} through the specimen. 
This is formally the transpose of the multislice operation defined in Eq.~(\ref{eq:MSop}), which is represented by a superscript $^T$ in the above algorithm description, and is given by
\begin{align} \label{eq:MSTop}
   \mathcal{M}^T(\Delta z)\equiv \hat{\mathcal{F}}_{\vect{\mathrm{r}\rightarrow\mathrm{h}}}e^{i\sigma V(\vect{r})\Delta z}\hat{\mathcal{F}}^{-1}_{\vect{\mathrm{g}\rightarrow\mathrm{r}}}e^{-i\pi\lambda \Delta zg^2\delta_{\vect{gh}}}\,.
\end{align}

The PRISM STEM-EELS algorithm requires $n_z*[N_1+N_2(n_z+1)]$ multislice operations where $N_1$ is the number of rows in matrix $\mathcal{S}_1$, which correspond reciprocal space points in the illumination, and $N_2$ is the number of columns in matrix $\mathcal{S}_2$, which correspond to reciprocal space points in the EELS detector. 
Since the PRISM STEM-EELS simulation algorithm is typically more economical with the required number of multislice iterations, the matrix multiplication step, evaluating Eq.~(\ref{eq:Hn0op}), tends to be the rate limiting step.
This means that although the calculation time is technically quadratic in $n_z$, in many cases run time scaling is instead predominantly determined by the number of unique transitions. 
If we further assume that the number of transitions in each slice is roughly constant  then the scaling will be roughly linear with the number of slices $n_z$.
In the upcoming Sec.~\ref{sec:inv_multislice}, we will introduce an approximation that makes the scaling truly linear with $n_z$.

Comparing the run time of the PRISM STEM-EELS algorithm to the conventional multislice approach requires accounting for the relative speed of FFTs to array multiplication.
To explore this question we make the approximation that the computation time of a two-dimensional FFT can be parametrized as $T_{\mathrm{FFT}}=A N^2\log N$ and that of an array multiplication (as used in the multislice operation) can be parametrized as $T_{\mathrm{mult}}=B N^2$ and that of an array multiplication and summation step (as would occur in a step of  a matrix multiplication where a single row and column are multiplied and added to the final result) can be parametrized $T_{\mathrm{addmult}}=C N^2$.
\begin{figure}
    \centering
    \includegraphics[width=\columnwidth]{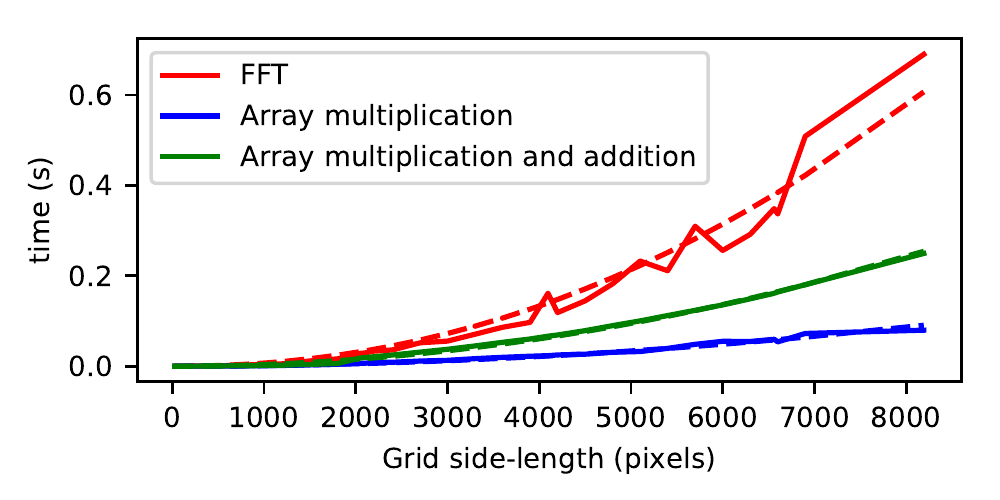}
    \caption{Scaling of FFT, array multiplication and array multiplication and addition operations in Matlab with the side length (pixels) of the square grid. The fitted scalings for these operations are plotted with dashed lines and the coefficients of the fit are given in the text. \label{fig:Matlabtime}}
\end{figure}
Here $A$, $B$ and $C$ are constants and $N$ is the size of the (square) grid using in computation. 
Measurements of $T_{\mathrm{FFT}}$, $T_{\mathrm{mult}}$ and $T_{\mathrm{addmult}}$ for different values of $N$ for Matlab are plotted in Fig.~\ref{fig:Matlabtime} for which values of $A=1.0\E{-9}$, $B=9.0\E{-9}$ and $C=3.8\E{-9}$ were fitted (fitted functions are plotted with dashed lines).
However, for the most accurate estimates of the runtime of calculations at a specific pixel grid size $N$, in the examples provided in this paper we will simply measure $T_{\mathrm{FFT}}$, $T_{\mathrm{mult}}$ and $T_{\mathrm{addmult}}$ at that pixel grid size $N$.
We also note that the platform and compiler can impact the relative speeds of different algorithms.

We assume a simulation object measuring $L\times L$ in the plane perpendicular to beam propagation and of thickness $z$, which we divide  into  $n_z=z/\Delta z$ slice. This object  would require a STEM scan with a Nyquist sampling of $(4L\alpha)^2$ probe positions, where $\alpha$ is  the probe forming aperture in units of inverse length~\cite{dwyer2010simulation}. 
Assuming a constant number of ionization states at each slice, not true in general as atoms of the element of interest might be not be uniformly dispersed throughout the sample, but a useful approximation for the current timing estimates, the computation time for the conventional multislice approach will be given by
\begin{align}\nonumber
    T_{MS} \approx 2(4L\alpha)^2 (n_\text{states}n_z(n_z-1)/2+n_z)&\\
                      (A N^2\log N + B N^2 )& \,. \label{eq:TMS}
\end{align}

The speed of the PRISM STEM-EELS calculation will depend on the size of the matrices used. For matrix $S_1$, the number of rows will be depend on the sampling of the probe forming aperture function $A(\vect{q})$ in Eq.~(\ref{eq:illum}) which covers a reciprocal space area of $\pi \alpha^2$ \AA$^{-2}$. 
A simulation cell size of $L$ implies a natural reciprocal space sampling of $L$ pixels per unit of inverse length for the illumination~\cite{kirkland2010advanced} though this can be further reduced to $L/f$ by using a PRISM interpolation factor of $f$ -- an optimization described in detail in Ref.~\cite{ophus2017fast}. 
The number of rows in the scattering matrix will then be given by $\pi \alpha^2L^2/f^2$. 
By similar reasoning, the second scattering $S_2$ matrix, which propagates the inelastically scattered electrons from the plane of ionization to the EELS aperture, will have a number of columns equal to $\pi\beta^2L^2/f^2$ where $\beta$ is the diffraction space size of the EELS aperture. 
The time required for multislice iterations necessary for the PRISM algorithm is therefore given by
\begin{align} \nonumber
    T_{\text{PRISM},\text{multislice}}=2(\pi \alpha^2 + 2*(n_z-1)\pi \beta^2)L^2/f^2&\\
    \cdot n_z (A N^2\log N + B N^2 )&\,. \label{eq:TPRISM1}
\end{align}
The number of columns in $S_1$ and the number of rows in $S_2$ will be given by the square of half total number of pixels in the simulation cell $(N/2)^2$.
These matrices need only have a side-length of $N/2$ since the output from a multislice calculation is bandwidth limited either to either 1/2 (as in the implementation used for this investigation) or 2/3 of the total array size and only spatial frequencies within this band-limit need be kept in the scattering matrix. 
For more detail on the need for this bandwidth limiting approach the reader is referred to Sec.~6.8 of Ref.~\cite{kirkland2010advanced}. 
With reference to the STEM-EELS simulation using scattering matrices we must sum  the computation times of the multislice iterations and the matrix multiplications,
\begin{align} \nonumber
    T_{\text{PRISM}} \approx B(N/2)^2\pi\alpha^2L^2/f^2&+C (N/2)^2
\pi^2\alpha^2\beta^2L^4/f^6\\ \label{eq:TPRISM2}
&+4C\pi^2\alpha^4\beta^2L^6/f^6
\end{align}

Where the first term is the time required to calculate the matrix multiplication of $\mathcal{S}_1\mathcal{H}_{n0}$ in Eq.~(\ref{eq:Hn0op}), the second term the time required to do the second matrix multiplication $(\mathcal{S}_1\mathcal{H}_{n0})\mathcal{S}_2$ in Eq.~(\ref{eq:Hn0op}) and the third term the time required to do the matrix multiplication $\mathcal{S}_n\boldsymbol{\Psi}$ for each probe ${\psi}$ in the STEM raster in Eq.~(\ref{eq:SHn0}).
In the following section we discuss a MATLAB implementation of the STEM-EELS simulation algorithms discussed show that the above expressions, Eqs.~(\ref{eq:TMS}), (\ref{eq:TPRISM1}) and (\ref{eq:TPRISM2}),  are correct estimators of the runtime of these calculations.

Memory requirements for both algorithms also merit discussion. At a minimum the multislice algorithm requires only arrays containing the Fresnel free-space propagator, transmission function, probe and ionization transition potential, so 4 $N\times N$ complex valued arrays. If there is sufficient memory, as is the case for most simulations, the transmission functions for all the slices will also be stored rather than calculated on the fly -- an additional $n_z$ $N \times N$ complex arrays. The PRISM STEM-EELS algorithm adds the requirement that two scattering matrices be stored in memory, an additional $\pi \alpha^2/f^2 + \pi \beta^2/f^2$ arrays of size $N/2\times N/2$. By way of example we discuss the requirements for simulation of the nanoparticle performed in the upcoming Sec.~\ref{sec:FePt}. The object was sampled on a 1836$\times$1836 grid and partitioned depth wise into 45 slices. The minimum of 4 single-precision complex floating-point arrays for the conventional multislice approach would take up 107 MB of memory. Storing the transmission functions takes up a further 1.21 GB. The number of columns in the first scattering matrix $\mathcal{S}_1$ and the number of rows in the second scattering matrix $\mathcal{S}_2$ are both 325. Each of these S-matrix rows and columns has size equal to a  918$\times$918 grid so approximately 4.38 GB is required to store both matrices. This is a substantially greater amount of memory than is the case for the equivalent multislice calculation, though we note that with current technology most high-end graphics cards have 8 GB or greater of memory and this example calculation is for a larger simulation cell than has typically been attempted before.
\section{Results and discussion \label{sec:Implementation}}
\subsection{Implementation of a scattering matrix based STEM-EELS simulation method \label{sec:implementation}}

\begin{figure}
    \centering
    \includegraphics[width=\columnwidth]{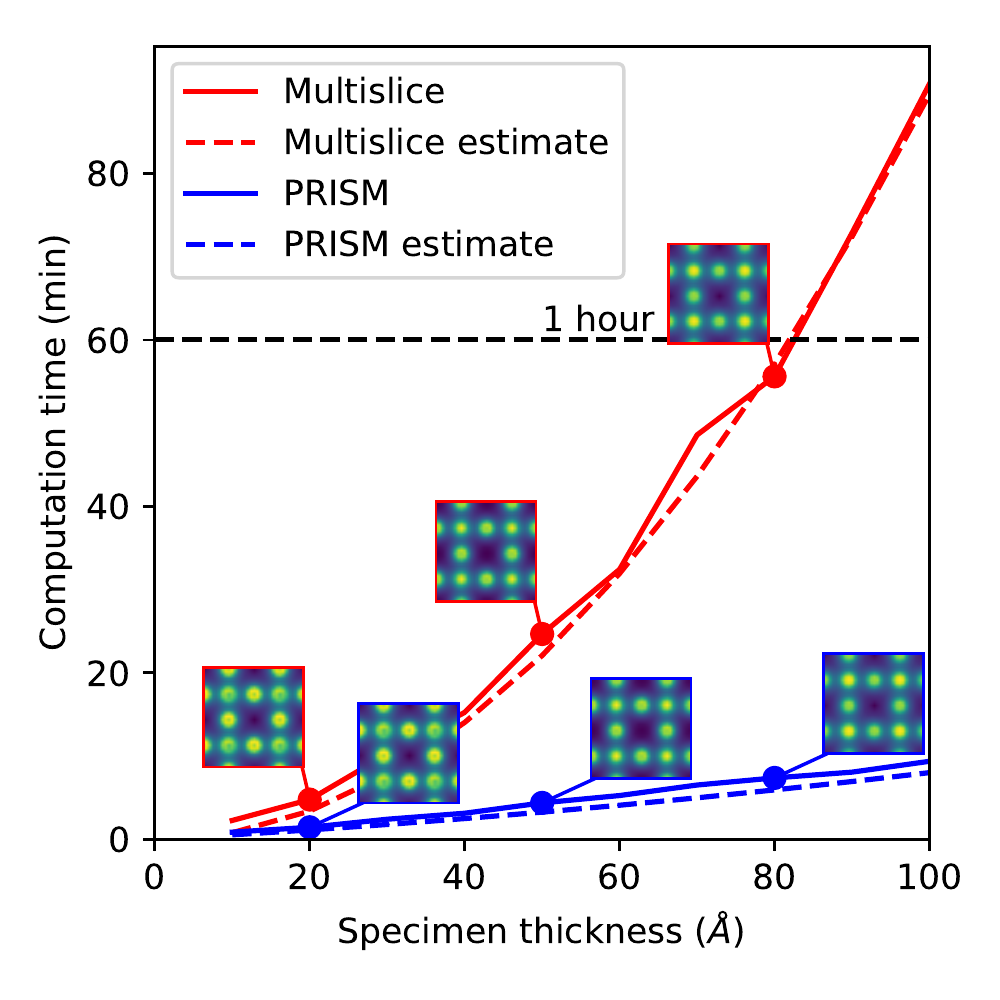}
    \caption{Scaling of computation time for standard multislice and PRISM approaches to the STEM-EELS image simulation with thickness for a SrTiO3 sample, calculation details are given in the text.}
    \label{fig:calcComparison}
\end{figure}

In this sectio nwe report results from a Matlab implementation of the conventional multislice algorithm and the new PRISM algorithm.
This implementation is included in the supplementary materials of this paper. 
It is included here to provide an accessible demonstration of the two algorithms introduced in the previous Sec.~{\ref{sec:theory}}; the code does not calculate the inelastic transition potentials $H_{n0}$ given by Eq.~(\ref{eq:Hn0op}) but uses a single calculated $H_{n0}$ outputted from the $\mu$STEM code~\cite{allen2015modelling}. 
As a test case we calculate STEM-EELS images of a single transition $(\ell=0,m_\ell=0)\rightarrow(\ell^\prime=1,m_\ell^\prime=1)$ for ionization of the O 1s orbital (the K-edge) using both the conventional multislice method (red solid line) and the new PRISM approach (blue solid line) for thicknesses between 10 \AA\ and 100 \AA.
A 2x2 tiling of the SrTiO3 unit cell (measuring 7.81 \AA\ $\times$ 7.81 \AA) specimen and a 160x160 pixel grid was used, parameters which are likely to result in somewhat unconverged calculations but result in faster runtimes for both algorithms and so allow rapid comparison of results.
For the purposes of simplifying comparison of conventional multislice and PRISM-EELS results thermal vibrations of the atoms where turned off for this calculation.
The actual timings for each algorithm are compared in Fig.~\ref{fig:calcComparison} with the relevant estimates from Eq.~(\ref{eq:TMS}) for the multislice case using measured $T_{\mathrm{FFT}}$, $T_{\mathrm{mult}}$ and $T_{\mathrm{addmult}}$ for a 160x160 pixel grid (red dashed line) and the sum of Eqs.~(\ref{eq:TPRISM1}) and (\ref{eq:TPRISM2}) for the PRISM case (blue dashed line) showing that these equations give reasonable estimates of computation time for these simulations. 
Images for the PRISM and conventional multislice calculations are shown for the thicknesses 10, 50 and 80 \AA. 
We compare differences in the images using the following normalized root mean sum of squares error percentage error metric,
\begin{align} \label{eq:nrmsse}
    \varepsilon =\frac{1}{100}\sqrt{\frac{(\sum I_{\text{MS}}(R) - I_{\text{PRISM}})^2}{(\sum I_{\text{MS}}(R))^2}}.
\end{align}
These are tabulated in Fig.~\ref{fig:errors}, both for the total image (Total error, $\varepsilon_T$) and in a 6 $\times$ 6 pixel window centered on the Ti-O column (site error $\varepsilon_S$).
Total error $\varepsilon_T$ is typically less than 0.01\% and site error $\varepsilon_S$ is between 0.001\% and 0.02\%. 
This low a discrepancy confirms that both the PRISM and conventional multislice approaches indeed encapsulate the same scattering physics.
The PRISM method is faster for all thicknesses and exhibits the pseudo-linear scaling predicted for this algorithm, whilst the scaling of the conventional multislice method is quadratic with thickness.

\begin{figure}
    \centering
\begin{tabular}{c|c|c} 
Thickness (\AA) &Total error ($\varepsilon_T$) & Site error ($\varepsilon_S$) \\ \hline
20 & 0.00207 \% & 0.00186 \%\\ 
50 & 0.00654 \% & 0.00855 \%\\ 
80 & 0.00779 \% & 0.00783 \%\\  
100 & 0.01040 \% & 0.01580 \%\\  
\end{tabular}
    \caption{Percentage error of PRISM calculation relative to multislice calculation for each thickness in Fig.~(~\ref{fig:calcComparison})}
    \label{fig:errors}
\end{figure}
\begin{figure}
    \centering
    \includegraphics[width=\columnwidth]{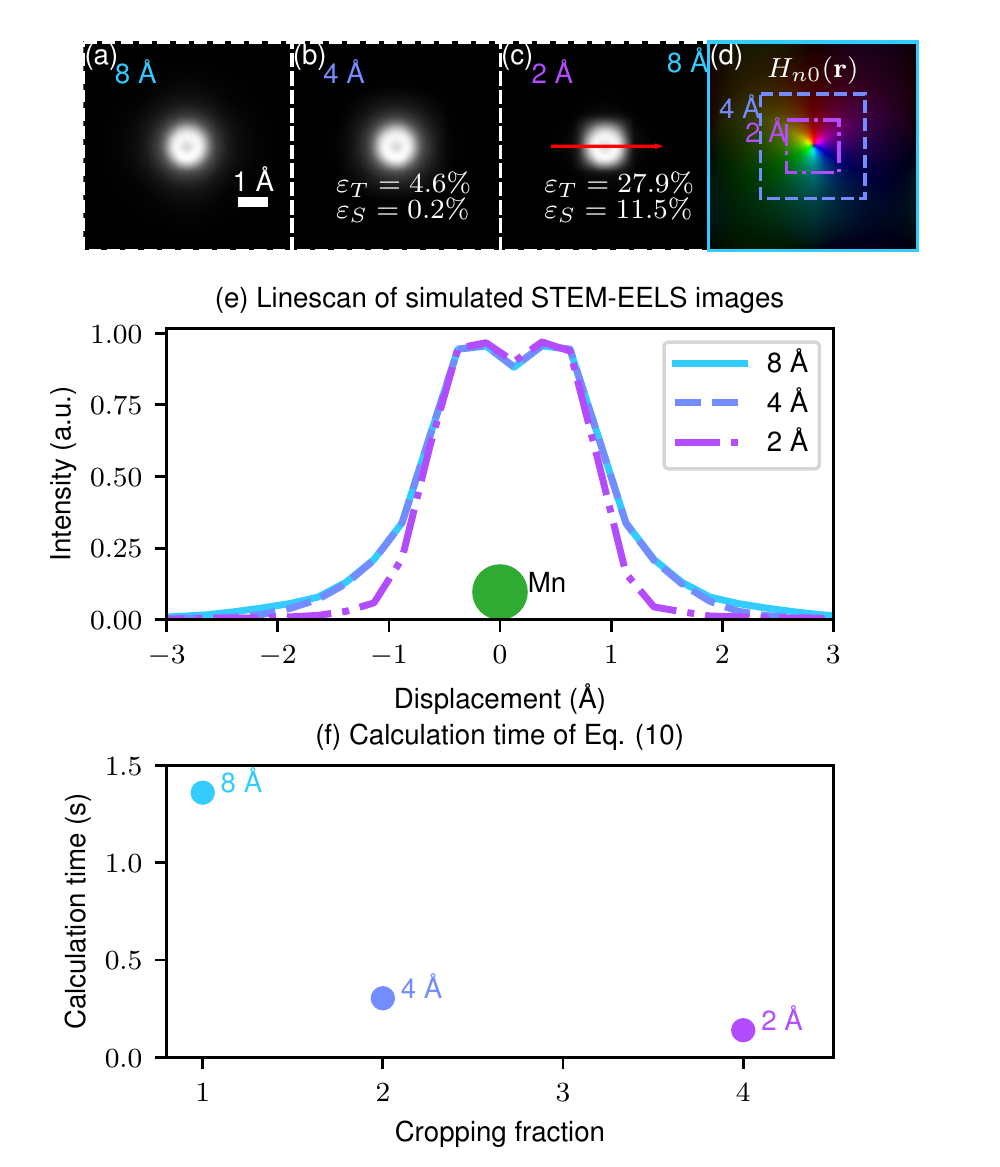}
    \caption{The STEM-EELS image for a single L shell transition for a dopant Mn atom within a 100 \AA\ thick SrTiO$_3$ crystal is calculated using different grids of side length (a) 8 \AA\ (b) 4 \AA\ and (c) 2 \AA\ to evaluate Eq.~(\ref{eq:SHn0}). The size of these windows is shown relative to the complex $H_{n0}(\vect{r})$ transition potential in the rightmost panel of (d). (e) A linescan through the centre of the image, indicated with a red arrow in (c), shows differences in the tails of the image. (f) The time required to calculate Eq.~(\ref{eq:SHn0}) for each of the images in (a)-(c).}
    \label{fig:cropfig}
\end{figure}
\subsection{Algorithm speed ups: calculating inelastic transitions on a cropped grid.}
Further speed-ups, with a modest penalty to accuracy, can be achieved by only evaluating the matrix multiplication in Eq.~(\ref{eq:SHn0}) in a fraction of the grid centered about the site location of the transition event $H_{n0}(\vect{r})$ rather than over the whole simulation grid. 
To demonstrate this optimization we simulate STEM-EELS images of a single Mn dopant occupying an Sr site midway through a 50 \AA\ thick SrTiO$_3$ crystal.
An L-shell transition for Mn is shown in Fig.~\ref{fig:cropfig} where Eq.~(\ref{eq:SHn0}) is (a) evaluated over the full grid (which has a side length of approximately 8 \AA), and in cropped regions with side-lengths measuring approximately (b) 4 \AA\ and (c) 2 \AA . 
The size of these windows relative to the transition potential $H_{n0}(\vect{r})$ is indicated in Fig.~\ref{fig:cropfig}(d).
Only the transitions $(\ell=1,m_\ell=1)\rightarrow(\ell^\prime=2,m_\ell^\prime=2)$ and $(\ell=1,m_\ell=-1)\rightarrow(\ell^\prime=2,m_\ell^\prime=-2)$ for energy loses 1 eV over the ionization thrheshhold where included for the purpose of this demonstration.
The percentage errors, relative to the image in Fig.~\ref{fig:cropfig}(a), are written on images Fig.~\ref{fig:cropfig}(b) and (c). 
Both the total image error $\varepsilon_T$ and error for the atomic site $\varepsilon_S$, which is evaluated only for a 1 \AA\ window around the Mn atomic position are indicated.
The site error $\varepsilon_S$ is the most relevant metric for our purposes since elemental concentration mapping would typically proceed by integrating the STEM-EELS signal in a window centered on the atomic site and then relating the result to a pre-computed look-up table.
A site error of 0.2\%, as in Fig.~\ref{fig:cropfig}(b) is likely acceptable, though a site error of 11.4\% as in Fig.~\ref{fig:cropfig}(b) is likely too high, suggesting the cropping window chosen was too aggressive.
Inspection of a linescan in Fig.~\ref{fig:cropfig}(e) shows that in the image from Fig.~\ref{fig:cropfig}(c) the long tails of the Mn transition potential $H_{n0}$ have been cropped out.
Figure ~\ref{fig:cropfig}(f) details the significant speed up benefits of this approach. 
The speed ups are quadratic which is consistent with the array multiplication scaling quadratically with grid pixel size, as the calculation time of Eq.~(\ref{eq:SHn0}) is observed to roughly quarter with each halving of the window in Fig.~\ref{fig:cropfig}(a).
For this particular transition potential a cropping box with side length of $4$ \AA\ gives the best balance between calculation time speed up and loss of accuracy.
\subsection{Algorithm speed ups: inverse multislice \label{sec:inv_multislice}}
In the PRISM STEM-EELS algorithm described in Sec.~\ref{sec:implementation}, the scattering matrix $\mathcal{S}_2$ which propagates the inelastically scattered electron wave $\psi_n$ from the depth at which  ionization  occurred to the exit surface is calculated from scratch for each thickness $i_z$.  
This is a duplication of work, since $\mathcal{S}_2$ was at some point calculated for all thicknesses in the initialization step $\mathcal{S}_2 = (\mathcal{M})^{T})^{n_z}$.
One approach would be to store $\mathcal{S}_2$ for each slice, which is not practical for most calculations given the size of the scattering matrix (a complex numbered array of size $\pi \alpha^2L^2/f^2\times N \times N$).
A second approach which sacrifices some accuracy at the expense of calculation time would be to perform the inverse of the multislice operation ($\mathcal{M}^{-1}$) to retreat the scattering matrix $\mathcal{S}_2$ a single slice.
The inverse multislice operation is defined
\begin{align} \label{eq:MSinvop}
   \mathcal{M}^{-1}(\Delta z)\equiv \hat{\mathcal{F}}_{\vect{\mathrm{h}\rightarrow\mathrm{r}}}e^{-i\sigma V(\vect{r})\Delta z}\hat{\mathcal{F}}_{\vect{\mathrm{r}\rightarrow\mathrm{g}}}e^{i\pi\lambda \Delta zg^2\delta_{\vect{gh}}}\,.
\end{align}
Relative to the forward multislice operation defined in Eq.~(\ref{eq:MSop}) the order of FFT, multiplication, inverse FFT and propagation steps has been reversed and the complex-conjugate of the transmission function and propagation operators is used instead. 
This can introduce some error since a forward multislice iteration can cause electrons to scatter to high angles outside the bandwidth limit of the calculation and these electrons will not be recovered with an inverse operation. 
For the simulation of the SrTiO$_3$  in Sec.~\ref{sec:implementation} we report the percentage errors, both total error $\varepsilon_T$  and site error $\varepsilon_S$, in Fig.~\ref{fig:inverse_ms}.
For this case, the errors relative to conventional multislice calculations are found to be around 1\%, which is small whilst the speed up is around 20\% relative to PRISM calculations for some of the thicker cases considered without the inverse multislice optimization.

\begin{figure}
    \centering
    \begin{tabular}{c|c|c|c}
    Thickness (\AA) &Total error ($\varepsilon_T$) & Site error ($\varepsilon_S$) & Speed up (\%) \\ \hline
    20 &  0.17 \% &  0.31 \% & 2.61 \%\\
    50 &  0.72 \% &  1.28 \% & 16.0 \%\\ 
    80 &  0.97 \% &  1.56 \% & 24.4 \%\\ 
    100 &  1.00 \% &  1.53 \% & 21.3 \%\\ \hline
    \end{tabular}
    \caption{Percentage errors in the PRISM STEM-EELS calculations with the inverse multislice optimization relative to conventional multislice calculations as well as the percentage speed up relative to PRISM STEM-EELS calculations without the inverse multislice optimization.}
    \label{fig:inverse_ms}
\end{figure}
\begin{figure*}
    \centering
    \includegraphics[width=2\columnwidth]{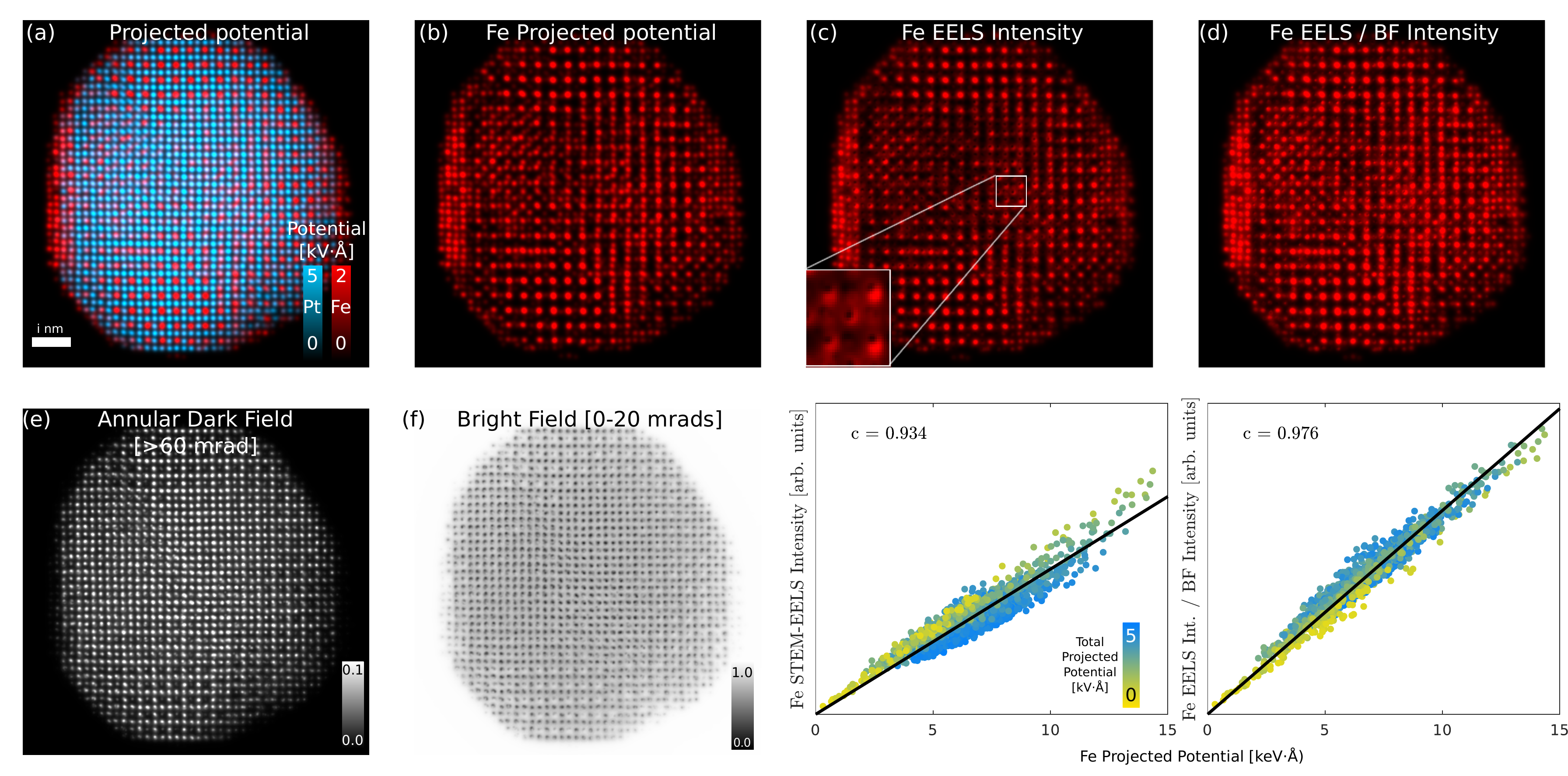}
    \caption{(a) The projected electrostatic potential for the FePt nanoparticle reconstructed in Ref.~\cite{yang2017deciphering} with Pt atoms indicated in blue and Fe atoms in red, the Fe projected potential is shown separately in (b). The STEM-EELS simulation of the structure in (c), which is corrected by dividing by an incoherent bright-field image in (d) as suggested in  Ref.~\cite{zhu2013towards}. The annular dark field and incoherent bright-field images are shown in (e) and (f). Figures (g) and (h) plot the relationship between Fe Projected potential and STEM-EELS intensity, averaged in 5$\times$ pixel windows, for (c) and (d) respectively.  \label{fig:FePt}}
\end{figure*}
\subsection{Calculations for heterogeneous nanometer scale objects \label{sec:FePt}}
In Sec.~\ref{sec:implementation} we showed that using the PRISM approach for STEM-EELS simulations resulted in computation times that scale more favourably with specimen thickness and that calculation time can be further reduced by evaluating the inelastic scattering cross section, Eq.~(\ref{eq:SHn0}), only in a fraction of the calculation grid centered on the transition of interest and using the inverse multislice optimization. 
In this section we demonstrate how these improvements, taken together, allow calculations of much larger objects than has previously been feasible, in this case a Fe-Pt nanoparticle that is approximately 80 \AA\ in diameter. 
The atomic coordinates for this nanoparticle were reconstructed from a STEM-HAADF tomography tilt series in Ref.~\cite{yang2017deciphering} and the object contains 6,569 Fe atoms and 16,627 Pt atoms. 
The projected potential of the nanoparticle is plotted in Fig.~\ref{fig:FePt}(a) with Pt potential shown in blue and Fe atoms shown red.
The potential resulting from only the Fe atoms is plotted seperately in Fig.~\ref{fig:FePt}(b).
Both potentials have been convolved with a Gaussian function ($\sigma$ = 2 pixels) to make viewing easier.

We consider the $(\ell=1,m_\ell=0)\rightarrow(\ell^\prime=2,m_\ell^\prime=1)$, $(\ell=1,m_\ell=0)\rightarrow(\ell^\prime=2,m_\ell^\prime=-1)$, $(\ell=1,m_\ell=1)\rightarrow(\ell^\prime=2,m_\ell^\prime=2)$ and $(\ell=1,m_\ell=-1)\rightarrow(\ell^\prime=2,m_\ell^\prime=-2)$ transitions which between them account for just over 90\% of total transitions for energy loses 1 eV over the ionization threshold Fe L edge. For the PRISM-EELS calculation, we use a PRISM interpolation factor of 9, such that each individual probe will be effectively calculated on a 10 \AA\ $\times$ 10 \AA\ grid, a probe step of 0.246 \AA\ (nyquist sampling)  and evaluate the inelastic scattering cross section [Eq.~(\ref{eq:SHn0})] on a 4 \AA\ $\times$ 4 \AA\ grid (which was seen to give the best trade-off between calculation speed and accuracy in Fig.~\ref{fig:cropfig}). 
Then, Eq.~(\ref{eq:TPRISM1}) and Eq.~(\ref{eq:TPRISM2}) estimate the run-time of such a calculation to be 2 days.
The results of this PRISM STEM-EELS simulation, which in reality took 16 hours, are shown in Fig.~\ref{fig:FePt}(c). We also estimate  the computation time of a conventional multislice simulation using Eq.~(\ref{eq:TMS}) for the full nanoparticle with the same probe step size and 2\AA\ slices along the beam direction.  If further approximations are made to use only 1/9 of the grid for probe propagation and only evaluating transitions within 4\AA\ of the probe (i.e. those that we deem to have a reasonable chance of being excited), multislice simulation could be run in about 87 days for a single frozen phonon pass \footnote{A full multislice calculation without these additional approximations is estimated to take 170 years for one frozen phonon pass.}. 

The simulated image in Fig.~\ref{fig:FePt}(c) is qualitatively similar to the Fe potential Fig.~\ref{fig:FePt}(b) but artifacts due to strong scattering of the beam are evident. In particular, the center of the nanoparticle has a lower intensity than would be expected from the density of Fe, due to the strong high-angle scattering of the electron beam from the heavier Pt atoms.
Close inspection of many of the Fe columns in this region reveals small regions of lower intensity, giving rise to a ``donut'' or ``volcano'' structure which is evident in the zoomed region shown in the bottom left hand corner of Fig.~\ref{fig:FePt}(c). 
These features result from inelastic scattering of beam electrons by the heavier Pt atoms, and to a smaller extent by the Fe atoms, to high angles outside the acceptance angle of the EELS detector when the beam is scanned atop an atomic column. 
Even if these scattered electrons do cause the ionization of an Fe atom they are unlikely to finally contribute toward the final STEM-EELS image.
A detailed explanation of this phenomenon can be found in Ref.~\cite{zhu2013towards} along with a strategy to obtain an image more amenable to direct interpretation: dividing by a simultaneously recorded STEM incoherent bright field image with detector of equal angular extent to the EELS aperture.
This image is displayed in Fig.~\ref{fig:FePt}(d) which is indeed observed to be a more faithful representation of the projected electrostatic potential due to Fe in Fig.~\ref{fig:FePt}(b).
The STEM annular dark field image, formed with a detector of inner angle 60 mrad and the incoherent bright-field image are shown in Figs.~\ref{fig:FePt}(e) and (f) for reference.

STEM-EELS simulations of objects of this size allow statistical analysis of STEM-EELS images as shown in Fig.~\ref{fig:FePt}(g), which plots the integrated potential just for the Fe atoms (a proxy for the projected Fe density) against the STEM-EELS intensity in for each 5$\times$5 pixel region in the image.
The colour of each point in the scatter plot corresponds to the total projected potential from \emph{both} Fe and Pt with reference to the colorbar in the figure.
The relationship between STEM-EELS intensity and projected Fe density is approximately linear.
However there are noticeable systematic deviations with points falling below the trendline tending to be those with a higher total projected potential -- a clear demonstration of how strong elastic and inelastic scattering of the beam complicates direct interpretation of the STEM-EELS maps.
These systematic errors are mostly remedied by the division of the incoherent bright-field image, which gives an improvement in the Pearson correlation score from 0.935 to 0.976. This correlation value measures the quality of the fitted trendline, showing that there is less systematic deviation from the linear relationship of Fe density and EELS intensity after applying the BF correction step. This insight demonstrates how our faster STEM-EELS algorithm, by virtue of its better scalability to larger simulation grids, can give a valuable insights into interpretation of heterogenous nanoscale STEM-EELS maps.

%
\section{conclusion}
We have developed a new algorithm for simulating STEM-EELS results that economises on the number of multislice iterations required. This algorithm should run faster in general as the calculation time typically scales linearly with specimen thickness, whilst the conventional algorithm scales quadratically with specimen thickness. We have shown that with a some penalty to accuracy, even faster calculation times are possible. Finally, we have also shown that our algorithm can be used to simulate larger nanoscale objects than was previously the case.

\section{acknowledgments}
The authors thank Les Allen and Scott Findlay for helpful discussions and proofreading during the preparation of this manuscript. Work at the Molecular Foundry was supported by the Office of Science, Office of Basic Energy Sciences, of the U.S. Department of Energy under Contract No. DE-AC02-05CH11231. All authors acknowledge additional support from the U.S. Department of Energy Early Career Research Program.

\section{Bibliography}

\end{document}